\documentclass[useAMS,usenatbib]{mn2e}
\usepackage{epsfig}

\newcommand{\PSbox}[3]{\mbox{\rule{0in}{#3}\includegraphics{#1}\hspace{#2}}}

\newcommand{\Ng}{{N}_{g}}
\newcommand{\Nobs}{{N}_{\rm obs}}
\newcommand{\Ntrue}{{N}_{\rm true}}
\newcommand{\Npeak}{{N}_{\rm peak}}
\newcommand{\Mmin}{{M}_{\rm min}}
\newcommand{\Mm}{{M}_{\rm min}}
\newcommand{\Mn}{{M}_{1}} 
\newcommand{\ncp}{{n}_{\rm cp}} 
\newcommand{\hmpc}{${h^{-1}}${\rm Mpc~}}

\newcommand{\hMsun}{{h^{-1}}{\rm M}_{\solar}}
\newcommand{\solar}{\ifmmode_{\mathord\odot}\;\else$_{\mathord\odot}\;$\fi}
\newcommand{\LCDM}{$\Lambda$CDM}

\newcommand{\beq}{\begin{equation}}
\newcommand{\eeq}{\end{equation}}

\def\ltsima{$\; \buildrel < \over \sim \;$}
\def\lsim{\lower.5ex\hbox{\ltsima}}

\begin{document}

\title[Galaxy Halo Occupation at High Redshift]
  {GALAXY HALO OCCUPATION AT HIGH REDSHIFT}
\author[Bullock, Wechsler, \& Somerville]
  {James S. Bullock$^1$, Risa H. Wechsler$^2$, Rachel S. Somerville$^3$\\
$^1$Department of Astronomy, The Ohio State University,
    140 W. 18th Ave, Columbus, OH 43210-1173; james@astronomy.ohio-state.edu\\
$^2$Physics Department, University of California, Santa
Cruz, CA 95064; risa@physics.ucsc.edu \\
$^3$Institute of Astronomy, Cambridge University, 
Madingley Rd., Cambridge, CB3 OHA, UK; rachel@ast.cam.ac.uk}

\maketitle

\begin{abstract}

We discuss how current and future data on the clustering and number
density of $z\sim3$ Lyman-break galaxies (LBGs) can be used to
constrain their relationship to dark matter haloes.  We explore a
three-parameter model in which the number of LBGs per dark halo scales
like a power-law in the halo mass: $N(M) = (M/\Mn)^S$ for $M>\Mmin$.
Here, $\Mmin$ is the minimum mass halo that can host an LBG,
$\Mn$ is a normalization parameter, associated with the mass above
which haloes host more than one observed LBG, and $S$ determines 
the strength of
the mass dependence.  We show how these three parameters are
constrained by three observable properties of LBGs: the number
density, the large-scale bias, and the fraction of objects in close
pairs.  Given these three quantities, the three unknown model
parameters may be estimated analytically, allowing a full exploration
of parameter space.  As an example, we assume a $\Lambda$CDM cosmology
and consider the observed properties of a recent sample of
spectroscopically confirmed LBGs.  We find that the favored range for
our model parameters is $\Mmin \simeq (0.4-8)\times 10^{10} \hMsun$,
$M_1 \simeq (6-10)\times10^{12} \hMsun$, and $0.9 \la S \la 1.1$.
The preferred region in $\Mmin$ expands by an order of magnitude
and slightly shallower slopes are acceptable
if the allowed range of $b_g$ is permitted to span all recent
observational estimates.
We
also discuss how the observed clustering of LBGs as a function of
luminosity can be used to constrain halo occupation, although due to
current observational uncertainties we are unable to reach any strong
conclusions.  Our methods and results can be used to constrain more
realistic models that aim to derive the occupation function $N(M)$
from first principles, and offer insight into how basic physical
properties affect the observed properties of LBGs.
\end{abstract}


\section{Introduction}
The Lyman-break color selection technique has made possible the
compilation of a large, fairly complete sample of $z\sim 3$ galaxies
(\citealt{steidel:98}; \citealt{adel:98}, hereafter
A98; \citealt{adel:01th}). This sample provides robust estimates of the
number densities and clustering properties of bright, high-redshift
galaxies, which can lead to invaluable constraints on models for the
evolution of structure in the Universe and high-z galaxy formation
\citep[A98;][]{giav:98,steidel:99,adel:00,giav:01,porciani:01}.

In  the CDM framework,  given a  power spectrum  and a  cosmology, the
number densities and clustering properties  of dark matter haloes  can
be readily  estimated at any redshift,  either by analytic  methods or
N-body simulations.  Relating     galaxies to these  dark  haloes   is
significantly  more  challenging,  since  the  way in  which  galaxies
populate haloes, both in number  and in luminosity, depends on aspects
of galaxy formation that are as of yet poorly  understood, such as the
efficiency of star formation and feedback processes (see, e.g.,
\citealt{sp:99}; \citealt*{spf:00}; \citealt{wech:01}, hereafter W01).
However, once the cosmological model is specified, the observed
clustering properties of galaxies can potentially be used to constrain
the nature of galaxy assembly.

The   current  observational estimates   of   the number  density  and
large-scale clustering  amplitude,    or  bias,  for these    $z\sim3$
Lyman-break galaxies (LBGs) are  reasonably consistent with a model in
which there   is  one galaxy  in  each  halo more  massive   than some
threshold                  (\citealt{mofuku:96};                  A98;
\citealt{wech:98,jing:98,bagla:98,coles:98,moscar:98,arnouts:99};
W01).  In  more  detailed models  of  galaxy  formation, however,  the
association between  haloes and galaxies  is  not expected to be  this
simple.  For example, even if the most luminous high-redshift galaxies
are quiescently star forming objects  that reside in massive haloes at
$z\sim3$, we    expect  that these   haloes will  contain substructure
comprised of haloes  that formed at  earlier epochs and have merged to
become      subhaloes     of     more     massive      hosts    (e.g.,
\citealt{kgkk:99,moore:99,springel:99,bullock:00};   W01).    If   the
high-redshift  galaxies  are  merger-triggered   starbursts, again  we
expect  to  find multiple galaxies  per halo,  perhaps  with different
occupation statistics than would be  predicted in a scenario dominated
by  quiescent  star  formation (\citealt{kolatt:99,bullock:99,spf:00};
W01).  It is therefore useful  to examine a  more general scenario for
populating haloes with galaxies, and  to explore ways of  constraining
the halo-galaxy relation directly.

In this  paper, we  will   focus on   how  the number  densities   and
clustering properties  of LBGs can be used  to constrain  the $z\sim3$
galaxy halo occupation function, $\Ng(M)$, which describes the typical
number of observed galaxies within a halo of mass $M$.  In addition to
using  the number  density  and large  scale  clustering amplitude  of
high-$z$ galaxies  to constrain the model,  we make use of a statistic
which reflects the small-scale clustering, the fraction of galaxies in
close pairs over  narrow redshift bins  (the close pair fraction).  As
an example of how  this can  be applied, we   use the number  density,
bias, and   close pair  statistics  derived  from  802 LBGs   from the
spectroscopically-confirmed sample of
\citet{adel:01th} to derive constraints on the general nature of the
halo occupation function.  Our framework is also applied to predict
clustering trends as a function of luminosity, and should prove useful
for interpreting future observations of high-$z$ galaxies.

In some respects, this work extends that of \citet{wech:00} and W01,
in which we used semi-analytic models of galaxy formation to predict
$\Ng(M)$ and then calculated the clustering properties of LBGs using
N-body simulations.  Here, we seek to constrain the halo occupation
function directly, using analytic approximations. We adopt a simple
functional form for the number of galaxies per halo as a function of
halo mass:
\begin{equation}
\Ng(M;\Mn, \Mmin,S)=\left(\frac{M}{\Mn}\right)^{S}; \quad  M \geq \Mmin. 
\end{equation}
This relation, which is motivated by the more detailed semi-analytic
modelling mentioned above, has three free parameters: $\Mmin$, the
minimum mass halo capable of hosting an observable LBG; $\Mn$, a
normalization parameter, which may be interpreted as the critical mass
above which haloes typically host more than one observed galaxy; and
$S$, the slope of the relation.  In principle, any model of galaxy
formation that aims to explain LBG properties can predict the value of
each of these parameters (as long as the observations can be
reasonably well described as a power law over some mass range; if not,
the approach discussed here can easily be extended to more complicated
functional descriptions).  Derived constraints on $\Mmin, \Mn$ and $S$
can serve as constraints on more sophisticated models and ultimately
as a probe of the underlying physics of galaxy formation.

Similar approaches, focusing mainly on the clustering properties of
local galaxies, have been performed to explore the $z = 0$ halo
occupation function
\citep[e.g.][]{jingmb:98,benson:00,seljak:00,peacock:00,
scoc:01,benson:01,berlind:01,cooray:01}.
Our focus on halo occupation at high redshift is complementary to
these local explorations, since together they provide a potential
probe for the evolution of star formation and galaxy assembly.  The
expected clustering properties of galaxies (and dark matter) in this
type of model can, in principle, be determined continuously over all
relevant length scales using analytic methods similar to those
presented by e.g., \citet{scoc:01}. However, the existing
observational samples at $z\sim3$ are currently too small to obtain
accurate estimates of the full correlation function and its moments.
For this reason, we focus on two measures of the clustering amplitude,
one at scales larger than the size (virial radius) of typical dark
matter haloes, reflecting the clustering properties of individual
haloes, and one at small scales, reflecting mainly the statistics of
multiple galaxies within common dark matter haloes.

In the next section (\S \ref{sec:observations}) we summarize the
current observational determinations at high redshift ($z\sim3$) of
the three main quantities used in our investigation: the comoving
number density, $n_{g}$, the large scale bias $b_g$ (which may be
related to the correlation length $r_0$), and the number density of
close pairs, $n_{cp}$, which may also be expressed as the close pair
fraction, $f_{cp} =n_{cp}/n_{g} $.  In the following section (\S
\ref{sec:model}), we outline our approach for predicting these three
quantities using our halo occupation model and analytic approximations
for the clustering properties of dark matter haloes.  In \S
\ref{sec:results} we use the observed estimates for the three numbers
$n_g$, $b_g$, and $f_{cp}$ to place constraints on the three model
parameters $\Mn$, $\Mm$, and $S$.  In \S \ref{sec:lumseg} we use our
model to make predictions for clustering segregation with luminosity,
and discuss how current and future observations help place further
constraints on halo occupation models.  We reserve \S
\ref{sec:conclusions} for discussion and conclusions.  In all
calculations, we adopt a flat CDM model with a non-zero vacuum energy
and the following parameters: $\Omega_m = 0.3, \Omega_{\Lambda} = 0.7,
h=0.7, \sigma_8=0.9$, where $\sigma_8$ is the rms fluctuation on the
scale of $8h^{-1}$ Mpc, $h$ is the Hubble constant in units of $100 $
km s$^{-1} {\rm Mpc}^{-1}$, and $\Omega_m$ and $\Omega_{\Lambda}$ are
the density contributions of matter and the vacuum respectively in
units of the critical density.

\section{Observational Quantities and Associated Uncertainties}
\label{sec:observations}
We focus this investigation on a relatively large sample of
Lyman-break galaxies, selected from a ground-based catalog of $U_n$,
${\mathcal G}$ and ${\mathcal R}$ photometry, which is complete to
${\mathcal R} = 25.5$ 
(\citealt{steidel:98}; A98; \citealt{adel:01th}).
Spectroscopic followup has been
performed for a subset of the photometric candidates, leading to
successful redshift identifications for about 45 percent of the total
sample of photometrically selected LBGs. All of the galaxies with
spectroscopic identifications have redshifts in the range $2.2\leq z
\leq 3.8$ (median redshift $\bar{z}\sim 3$). K. Adelberger has kindly
provided us with the data for 802 spectroscopically confirmed LBGs,
which consist of the 500 galaxies in the sample described in A98, plus
302 additional galaxies. At the time of writing it comprises the
largest and most complete sample of this kind. We shall refer to this
as the A01 sample. Recent analyses of the clustering properties of
subsets of this data have been presented by
A98, \citet{giav:98}, \citet{adel:00}, \citet[][hereafter GD01]{giav:01}, 
and \citet{porciani:01},
and we shall also make use of these results.

We choose statistics that can be calculated reasonably robustly from
this sample, and which produce constraints on the three free
parameters of our model. The statistics that we shall consider are:
\begin{enumerate}
\item the comoving galaxy number density, $n_g$
\item the large-scale galaxy bias, $b_g$
\item the close pair fraction, $f_{cp}$
\end{enumerate}
Each of these statistics can be derived directly from the data with a
small number of additional assumptions. We discuss the definitions of
each of these quantities in more detail below.

\subsection{Number Density}
Consider a population of galaxies with a given magnitude limit and at
a given redshift, with a (comoving) volume density $n_{\rm true}$. Of
course, no observed sample of galaxies is perfectly complete, and so
the observed density $n_{\rm obs}$ differs from the underlying, true
density by a factor $p \equiv n_{\rm obs}/n_{\rm true}$. Here, the
observed number density is just the number of galaxies actually
observed per unit redshift and solid angle, $\Nobs$,
divided by $dV \equiv \frac{\rm{d}V}{\rm{d}A\rm{d}z}(z)$, the comoving
volume element per unit redshift and solid angle. In the case of the
Lyman-break galaxies, which are pre-selected by color, the observed
population may differ from the underlying one at a given magnitude
limit for several reasons. Galaxies may be missing from the sample
because of confusion blending with nearby sources, or because their
observed colors lie outside the selection window, either intrinsically
or due to scattering because of photometric errors. In addition,
spectroscopic follow-up is only attempted for some fraction of the
color-selected candidates, and not all of these are successfully
assigned redshifts, usually because of insufficient
signal-to-noise. At a given redshift, we can write the relationship
between the number of galaxies in the true and observed population as:
\begin{equation}
\Nobs = f_{\rm spec} f_{\rm phot} \Ntrue\, ,
\end{equation}
where $f_{\rm spec} \equiv N_{\rm spec}/N_{\rm phot}$ is the fraction
of photometric candidates with successful redshift identifications,
and $f_{\rm phot} \equiv N_{\rm phot}/N_{\rm true}$ is the fraction of
the underlying population selected by the color-color criteria. In
principle, both of these terms may depend on redshift. We can write
the incompleteness of the photometric sample as $f_{\rm phot} = f_{\rm
peak} \phi(z)$, where $\phi(z) = \Nobs(z)/\Npeak$ is the
peak-normalized selection function, which is just the observed
redshift distribution normalized by the value at the peak.

If we make the simplifying assumption that the probability of
obtaining a successful redshift $f_{\rm spec}$ does not depend on the
redshift itself over the relevant range, then we can write the overall
selection probability $p$ as the product of three parts:
\begin{equation}
p = f_{\rm spec} f_{\rm peak} \frac{V_{\rm eff}}{V_{\rm top}} .
\label{eq:p}
\end{equation}
Here, $V_{\rm top}$ is the volume per unit area integrated over the
redshift range, and $V_{\rm eff}$ is the selection function weighted
volume per unit area:
\begin{equation} 
V_{\rm eff} = \int_{0}^{\infty} \phi(z) dV(z) dz 
\end{equation}
The selection function $\phi(z)$ may be constructed from the measured
redshift distribution averaged over many fields.  This function is
roughly Gaussian, with a mean of $\bar{z}\sim 3$ and a width of
$\sigma_z \sim 0.24$ (see A98; \citealt{giav:98}). Thus two of the
components of $p$ are well constrained observationally: the fraction
$f_{\rm spec}$ is trivially determined by relating the number of
galaxies in the spectroscopic sample with the original number of
photometric candidates --- our sample of 802 galaxies was selected
from a population of 1781 photometric candidates, so in this case
$f_{\rm spec}=0.45$.  What is uncertain is to what degree the
spectroscopic sample is biased towards objects of brighter magnitudes
(GD01); Eqn. \ref{eq:p} assumed that $f_{\rm spec}$ was independent of
the magnitude of the candidate.  If there is a strong bias toward
brighter galaxies affecting the completeness as a function of
magnitude, then the effective $f_{\rm spec}$ and $p$ values would be
increased to get a lower value of $n_{\rm true}$ for this brighter
sample.  The contribution to $p$ due to the selection function is also
well-constrained observationally: integrating over the selection
function gives $V_{\rm eff}/V_{\rm top} = 0.52$. The factor $f_{\rm
peak}$ is the most uncertain, but is probably in the range 0.5 to
1.0. Taken together, favored values are in the range $p=0.1$--0.3. In
W01, we assumed a value of $p=0.14$.

We  calculate the observed number   density $n_g$ directly by dividing
the number of galaxies  in the A01 sample  by the volume of the region
subtended by  the total angular size  of the survey (nine 9 arcmin$^2$
fields, one 6.5 arcmin$^2$ field, and three $7\times14$ arcmin fields)
over  the redshift range $2.5 \leq  z \leq 3.5$.  The implied observed
number density is then $n_g  = 6.6\times10^{-4} h^{3}$Mpc$^{-3}$.  The
error on this  number from cosmic  variance should be small  ($\sim 3$
per cent, based on resampling a large volume N-body simulation), so we
neglect it for the purposes of  normalizing our models. Because of the
remaining uncertainty in  the value of  the selection probability $p$,
we will work  only with the  observed number density.  The constraints
that we obtain can  then be translated back  to the values relevant to
the underlying, intrinsic  population if and when  the value of $p$ is
determined.

\subsection{Bias}
\label{sec:bias}
We define the bias as the square root of the  ratio between the galaxy
correlation function and dark matter correlation function: $b_{g}
\equiv [\xi_{g}/\xi_{DM}]^{1/2}$. It should be noted that several
different definitions of bias are used in the literature, and are not
equivalent \citep[see, e.g.,][]{dekel:99,slsdkw:01}.
Therefore, caution should be used when comparing bias values given by
different authors. An additional complication is that the bias may be
a function of the spatial scale on which it is measured.  If we adopt
a cosmology and power spectrum, a definition of bias, and a spatial
scale, and if the correlation function of galaxies is well-represented
as a power law, $\xi = (r/r_0)^{-\gamma}$, then for any galaxy
population where $r_0$ and $\gamma$ are determined, we can translate
this to a bias value.

We wish  to define the  `large-scale' bias on  a  scale that is larger
than the size of individual  haloes, so that  it is mainly affected by
the clustering  properties of distinct  haloes  themselves, and is not
significantly affected by  halo exclusion  effects  or the  occupation
function within haloes. This  scale is approximately $\ga 1$--2 \hmpc\
for the relevant halo masses at $z=3$. However, we encounter a problem
in defining a sensible value  of the large  scale bias for the case of
LBGs at $z=3$. The correlation function of the dark matter at $z=3$ in
our  chosen   cosmological model\footnote{We    have  calculated   the
correlation function for the dark matter from the publically available
GIF simulation, described in  some detail in  W01.} cannot be well fit
by a single power law, rather it resembles a broken power-law with the
break occurring at about 1 \hmpc. The slope at scales smaller than 1
\hmpc\ is roughly $\gamma_{DM}=1.5$, and at larger scales the slope
changes to a shallower slope of about $\gamma_{DM}=1.2$. If the
correlation function of LBGs is really a power-law with a slope of
$1.5-1.6$, as indicated by observations
(\citealt{adel:00,porciani:01}), then this implies that the bias
on scales $\ga 1$ \hmpc\ is strongly scale-dependent, and does not
assymtote to a stable value on any reasonable scale. The implied bias
values range from $b\sim1.8$ at a scale of 8 \hmpc, which is the
largest radius where the correlation function of LBGs is
observationally determined, to $b\sim2.6$ at 1 \hmpc.  At no point is
the bias really constant over any significant range of
scales. However, we do not wish to deal explicitly with the scale
dependence of the bias and the detailed shape of the correlation
function here, as the current observational constraints do not warrant
such a detailed investigation.  We thus assume that both the halos and
dark matter have $\gamma=1.5$ on all scales, and can thus define the
bias as $b=(r_{0,DM}/r_{0,g})^{-1.5/2}$, with $r_{0, DM}=1.2$ \hmpc.
This approximation works best at scales of about 2 \hmpc, but we find
that the approximation for halo bias \citep{smt:01} that we discuss in
the following section, in combination with this assumption about the
dark matter correlation function, agrees with the halo correlation
function measured in simulations within about 10 per cent over all
relevant scales.
This is basically equivalent to defining the bias at a scale of
$\simeq 2$ \hmpc.

The observational estimates of $r_0$ and $\gamma$ for LBGs vary
somewhat depending on the sample and the technique used to obtain the
three-dimensional, real-space correlation function from the projected
or redshift space data. A summary of the observationally derived
correlation function parameters and implied bias values is given in
Table~\ref{tab:obscf}.  We do not have a measured bias value for the
full A01 sample, but this sample is quite close to that of the
\citet{adel:00} sample and we use those values to constrain our model.

\begin{table*}
\label{tab:obscf}
\caption{Observational correlation function parameters,
for several different samples and methods, assuming the same \LCDM\
cosmology used throughout our analysis. SPEC indicates a sample with
spectroscopic redshifts, PHOT to the ground-based samples of
photometric LBG candidates, and HDF to the deeper sample of $U_{300}$
drop-outs from the HDF North and South. HDF-N photo-z is the sample of
HDF North galaxies with photometric redshifts in the range $2.5 < z <
3.5$. All magnitude limits are given in the AB system, and are the
authors' stated completeness limits (note that the SPEC sample of A98
and GD01 are just subsamples of the Adelberger 2000 sample). CIC
refers to the counts-in-cells method and $w(\theta)$ to the inversion
of the angular correlation function; ang CIC refers to a
counts-in-cells method of measuring $w(\theta)$. Where $\gamma$ is
given in square brackets, this indicates that the value was assumed
rather than derived.  The bias values are calculated under the
assumption that $\gamma=1.5$ for both galaxies and dark matter, and
that $r_{\rm 0,DM}=1.2$.  While many of the samples have
considerable overlap, they do not necessarily consist of the same
galaxies.
}
\begin{center}
\begin{tabular}{ccccccc}
\hline
Sample & Method & magnitude limit & $r_0$ [comoving \hmpc] & $\gamma$ & reference & bias \\
\hline
SPEC  & CIC & ${\cal R}=25.5$ & $6 \pm 1$ & [1.8] & A98 & $3.3$\\
SPEC & $w(\theta)$ & ${\cal R}=25.5$ & $3.8 \pm 0.3$ & $1.61 \pm 0.15$ & Adelberger 2000 & $2.4$\\
SPEC & CIC & ${\cal R}=25.0$ & $5.0 \pm 0.7$ &  $2.0\pm 0.2$ & GD01 & 2.9\\
PHOT & $w(\theta)$ & ${\cal R}=25.5$ & $3.2 \pm 0.7$ & $2.0\pm 0.2$ & GD01 &2.1\\
PHOT & ang CIC &  ${\cal R}=25.5$ &  $4.1^{+1.0}_{-1.5}$ & $1.50^{+0.25}_{-0.5}$ & PG01 & 2.5\\
HDF & $w(\theta)$ & $V_{606}=27$ & $1.4 \pm 1.0$ & $2.2^{+0.3}_{-0.2}$ & GD01 & 1.1\\
HDF-N photo-z & $w(\theta)$ & $I_{814}=28.5$ & $2.78 \pm 0.68$ & [1.8] & Arnouts et al. 1999 & 1.9\\
\hline
\end{tabular}
\end{center}
\label{tab:cf}
\end{table*}

\subsection{Close Pairs}
The close pair count $\ncp$ describes the number of pairs of galaxies
within a fixed angular separation on the sky and within a fixed
separation in redshift $\pm \Delta z$.  The value of $\ncp$ provides a
useful probe of small-scale clustering and is especially sensitive to
halo occupation statistics.  If the angular separation is chosen to be
slightly larger than the typical angular size of a halo ($\sim 0.4$
\hmpc\ comoving, or $\sim 20$ arcsec for our cosmology), then the
close pair fraction can probe the number of objects within haloes
without being sensitive to the details of how galaxies are distributed
spatially within them. Here, we focus on this single angular scale,
which we find to be most useful in constraining our chosen model
parameters, although in principle of course a range of angular
separations could be investigated (as we did in W01).  In order to
best separate the effects of projection, it is useful to limit the
pair counts to those galaxies that are within a small redshift range
of each other; however, the resolution of the data is not sufficient
to completely remove projection effects.  Looking on different scales
may also help to distinguish which pairs are in the same halo.  We
thus calculate the close pair counts for several choices of the
redshift bin size for the A01 sample.  Defining the close pair
fraction as just the number of close pairs divided by the total number
of galaxies, $f_{cp} =n_{cp}/n_{g}$, we find $f_{cp}(20'') = 0.010 \pm
0.004$, $0.015 \pm 0.004$, and $0.022 \pm 0.005$ for redshift bins of
size $\Delta z = 0.005$, $0.010$, and $0.040$ respectively.  The
errors reflect $1-\sigma$ statistical uncertainties.  Note that there
does not seem to be a strong bias against selecting close pairs; the
small-scale spectroscopic pair counts with no redshift selection are
almost identical to those of the photometric sample (after taking
number density into account; see also W01).

\section{A General Model for Galaxy Clustering}
\label{sec:model}

In this section we present the analytic expressions used to predict
the three observables introduced above in
Section~\ref{sec:observations}: the number density of observed
galaxies, $n_g$, the large-scale galaxy bias, $b_g$, and the close
pair fraction, $f_{cp}$. In the expressions for derived quantities
that follow, we will suppress the $S,\Mn,$ and $\Mmin$, dependence ---
such a dependence should be assumed unless otherwise noted.

The comoving number density of galaxies is the integral over $dn/dM$,
the differential number density of dark matter haloes as a function of
halo mass $M$, weighted by the appropriate galaxy occupation function:
\begin{eqnarray}
\label{eqn:numdens}
n_{g} =  \int_{\Mmin}^{\infty} 
\frac{dn}{dM}(M)\Ng(M) dM.
\end{eqnarray}
For the halo mass function, we use the analytic expression developed
by \citet{st:99}, which agrees fairly well with the results of N-body
simulations (see e.g. \citealt{jenk:01,sigad:01,rw:01}):
\begin{equation}
\frac{dn}{dM} = - \frac{\bar{\rho}}{M} \frac{d \sigma}{d M}
\sqrt{\frac{a \nu^2}{w}} \left[ 1 + (a \nu^2)^{-q} \right] \exp[
\frac{-a \nu^2}{2}].
\end{equation} 
Here, $\sigma$ is the linear rms variance of the power spectrum on the
mass scale $M$ at redshift $z=3$ and $\nu \equiv \delta_c / \sigma$,
where $\delta_c \simeq 1.686$ is the critical overdensity value for
collapse.  The other parameters are $a=0.707, q = 0.3,$ and $w=
0.163$, which were chosen to match N-body simulations with the same
cosmology and power spectrum as the one we have assumed.

We determine the large-scale bias for galaxies by integrating the
expected bias of haloes as a function of mass $b_h(M)$, weighted by the
galaxy occupation function $\Ng$:
\begin{eqnarray} 
\label{eqn:bias}
b_g = 
\frac{1}{n_g} \int_{\Mmin}^{\infty} 
\frac{dn}{dM}(M) b_h(M) \Ng(M) dM. 
\end{eqnarray} 
For the halo bias $b_h$, we use the expression of \citet{smt:01}
based on ellipsoidal collapse:
\begin{eqnarray}
\label{eqn:halo_bias}
b_h(M) = 1 + 
   \frac{1}{\sqrt{a}\delta_c} \left[
   \sqrt{a}(a \nu^2) + \sqrt{a}b(a\nu^2)^{1-c} -   \right.\\  \nonumber
   \left.\frac{(a\nu^2)^{c}}{(a\nu^2)^{c} + b(1-c)(1-c/2)} \right],
\label{eqt:stb}
\end{eqnarray}
where $b=0.5$ and $c=0.6$.
Note that because the bias is unaffected by random sampling or the
overall normalization, $b_g$ is independent of $\Mn$ (and also any
uniform selection probability $p$).

The number density of galaxies in close pairs $\ncp$ can be written as
the contribution of two pieces:
\begin{equation}
\ncp =  \ncp^{h} + \ncp^{d}.
\label{eq:ncp}
\end{equation} 
The first piece, $\ncp^{h}$, is the number density of close pairs of
galaxies within the same halo, and the `distinct halo' piece,
$\ncp^{d}$, represents galaxy pairs coming from objects that do not
lie within the same halo, and are counted as close pairs because of
projection effects.

In order to calculate $\ncp^d$, we need the correlation function of
galaxies inhabiting distinct host haloes $\xi_{d}(r)$. On scales
larger than the typical halo size, $\xi_g$ will mirror the halo
correlation function: $\xi_g(r) = \xi_d(r) = b_h^2 \xi_{DM}$ (with
$b_h$ the halo bias calculated from Eqn.~\ref{eqn:halo_bias}).  We
expect this assumption to break down on small scales, near the scale
where the virial radii of the haloes begin to overlap, $d_h = 2 R_{\rm
v}$, where $d_h$ is the diameter (twice the virial radius) of the
average-mass halo under consideration (which implicitly depends on
$\Mmin$).  The fact that haloes are mutually exclusive in space
demands that the correlation function go to zero (and to $-1$)
at some scale below
$d_h$.  A simple assumption is $\xi_{d}(r)=0$ for $r < d_h$, and
remarkably, when we make this assumption, we reproduce the projected
close pair counts derived from the N-body simulations discussed in W01
to an accuracy of 5-20 per cent.  Although the true nature of $\xi_d$
is certainly more complicated, a calculation of this accuracy is
sufficient for the level of observational precision relevant to this
work, so we adopt this simple break-radius form for $\xi_d$ for the
rest of our analysis.

Given $\xi_g$ and the volume $V$ defined by the bin geometry, the
number of expected pairs is now straightforward to determine.  The
average number of pairs within $V$ is:
\begin{equation}
\left< {\rm N}_{\rm pairs} \right>_V = 0.5 \left< N(N-1) \right >_V.
\end{equation}
Making use of 
\begin{equation}
\sigma^2(N) = \left < N^2 \right > - \left <N \right >^2 = n_g^2 
\int_V \xi_{d}(r_{12}) dV_1 dV_2 + 
\left <N \right >,
\end{equation}
we obtain $2<{\rm N}_{\rm pairs}> = \sigma^2(N) - \left<N\right> + \left< N
\right>^2$.  The number density of pairs for galaxies in
distinct host haloes is therefore
\begin{equation}
n_{\rm cp}^{d} = 0.5 n_g^2\left(\frac{1}
{V}\int_V \xi_{d}(r_{12}) dV_1 dV_2 + V \right).
\label{eq:ncpd}
\end{equation} 
Here, $r_{12}$ is the  distance
between the   volume elements $dV_1$   and $dV_2$.   

The second piece of the close-pair expression, $\ncp^h$, is obtained
by integrating the expected pair counts, $\left< \Ng(\Ng -
1)\right>$, in haloes of a given mass over the halo mass function.
Later, we will explore how scatter in the halo occupation function
affects the close pair counts, but for now, we make the limiting
assumption of zero variance.  This implies $\left< \Ng(\Ng -
1)\right> = \Ng(M)[\Ng(M)-1]$, and gives
\begin{eqnarray}
\label{eqn:ncp}
\ncp^{h} = 0.5 \int_{M_*}^{\infty} \frac{dn}{dM} \Ng(M)[\Ng(M) - 1]
 dM. 
\end{eqnarray}
The lower limit of the integral is $M_*=$max($\Mmin,\Mn$).

This two-piece approximation for calculating the expected close pair
counts provides a clear and intuitive picture for what the close pairs
represent physically.  The  projected piece, $\ncp^d$,  depends mainly
on $\xi_g$ and   $n_g$,  so that   at  a fixed   bias, it   is  nearly
independent of how the halo occupation function varies with mass.  The
$\ncp^{h}$ piece, however, depends strongly on the occupation
function, and in particular on the slope $S$.  Note that the
close pair calculation neglects redshift space distortions.

We stress again that as we are dealing with a sample which has an
uncertain relationship to the intrinsic underlying population, all of
the above definitions refer to the numbers of objects \emph{that would
be included in our observational sample}. For example, we explicitly
define the occupation function, $\Ng$, to be the number of {\em
observed} LBGs per halo, rather than the actual number of galaxies per
halo.  In this language, the number of galaxies that actually exist
per halo will be $p^{-1}\Ng$.  This uncertainty in normalization
translates to an uncertainty in the `intrinsic' value of $\Mn$ via
$\Mn^i = p^{1/S}{\Mn}$, but does not affect the other estimates.
The special case of one-galaxy-per-halo ($S=0.0$) is a bit
different.  For models of this kind, $\Mn$ is undefined, and $p$ must
be defined explicitly.  However, in this case, the model will still
contain the same number of parameters, with $\Mn$ replaced by $p$.

The model outlined above is general in the sense that it could be
applied to any population of galaxies whose halo occupation function
is well approximated by a power law, over scales where the assumption
of linear, scale-free) bias is sensible. Here, we proceed to apply it
to the single example of Lyman-break galaxies in the \LCDM\ cosmology
specified above.

\begin{figure}
\PSbox{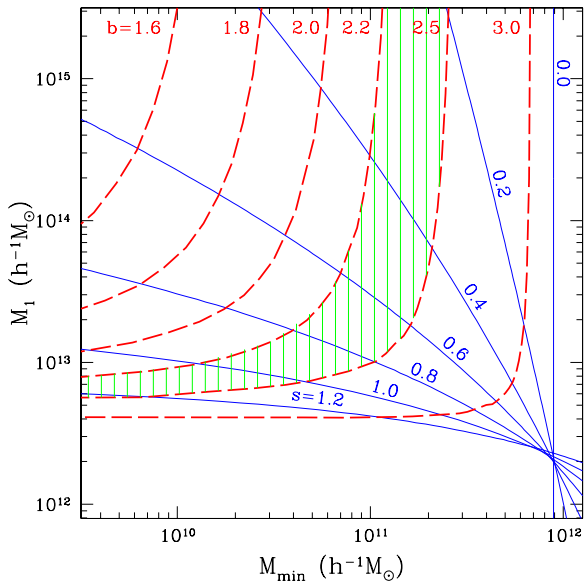  hoffset=-10 voffset=-65 hscale=40 vscale = 40}
{74mm}{74mm}
\caption[Relation between model parameters $\Mn$, $\Mmin$, $S$,and
bias $b.$] {The critical mass $\Mn$ for hosting more than one
observed galaxy as a function of the minimum mass for hosting any
observed galaxy, for different values of the halo occupation slope and
galaxy bias.  The observed number density is fixed to the observed
value for the A01 sample of $z\sim3$ Lyman-break galaxies (see text). 
Thin solid
lines correspond to fixed values of S; dashed lines correspond to
constant bias values. 
The shaded band indicates
the allowed region for $b_g = 2.2-2.5$, the range of bias values 
favored by the Adelberger (2000) analysis.
 }
\label{fig:m0-mmin}
\end{figure}

\section{Constraints From Lyman-Break Galaxies}
\label{sec:results}
In this section we will use the observational data derived from the
$z\sim3$ sample of Lyman-break galaxies summarized in
Section~\ref{sec:observations} to constrain the halo occupation
function for these objects. Fig. \ref{fig:m0-mmin} shows the
relation between the model $\Mn$ and $\Mmin$ values obtained by
inverting equations \ref{eqn:numdens} and \ref{eqn:bias} for given
values of the number density $n_g$, bias $b_g$, and occupation
function slope $S$. In all cases, we fix the number density to the
observed value of $n_g$ given in Section~\ref{sec:observations}.  The
thin solid lines show the relation for different values of the
occupation function slope $S$.  As $\Mm$ decreases, galaxies may
inhabit smaller mass haloes, which are more numerous.  In order to
maintain the observed number density, the mass $\Mn$ at which a host
halo contains one galaxy must increase correspondingly.  The change in
$\Mn$ as a function of $\Mm$ is steeper for smaller values of $S$.

If instead of fixing $S$, we fix the bias $b_g$ (as well as the number
density), we obtain the dashed lines in Fig. \ref{fig:m0-mmin}.  By
comparing the dashed and solid lines, it is evident that for fixed
values of $\Mmin$, the bias is an increasing function of the slope
$S$.  This is because, for fixed $\Mmin$, and for larger values of
$S$, more galaxies reside in larger mass hosts.  It is also evident
that the galaxy bias is a stronger function of $\Mmin$ for high-$S$
models.

\begin{figure*}
\centerline{\psfig{file=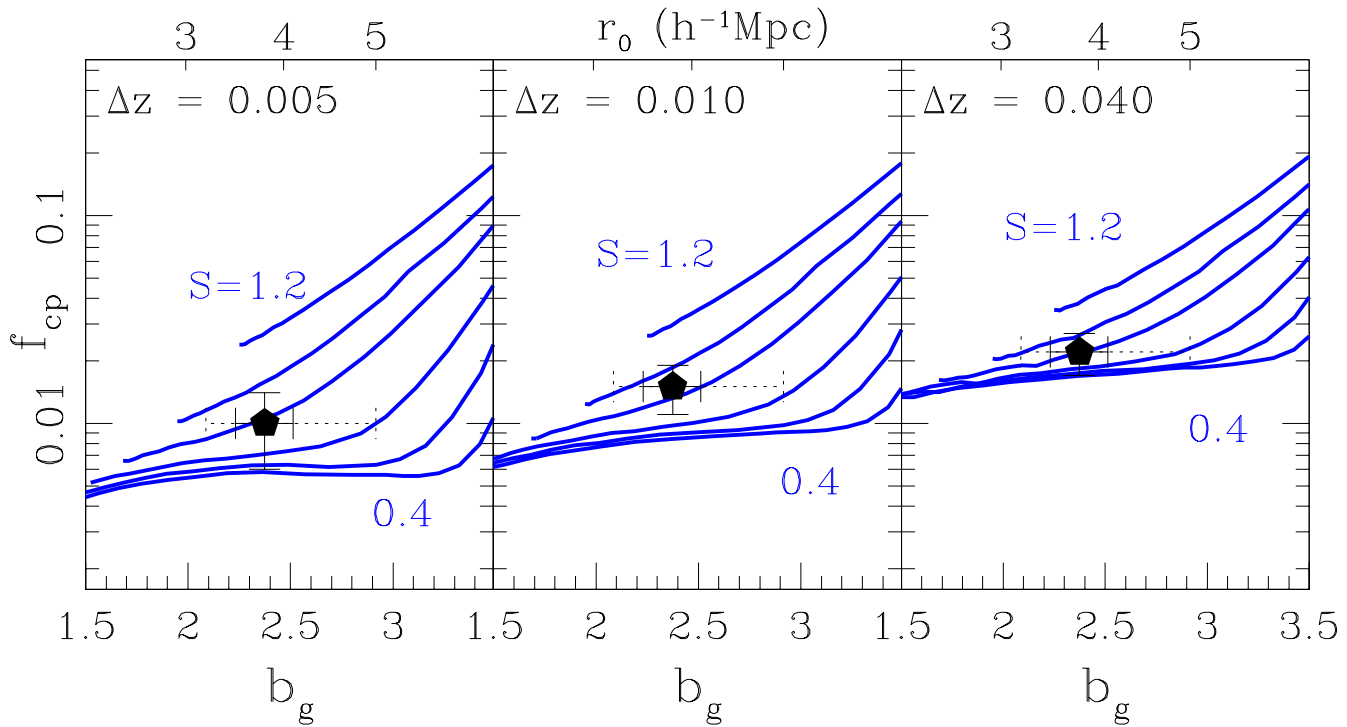,width=7in}}
\caption[Pair fraction vs. galaxy bias] {The fraction of pairs within
an angular separation of $20''$ for three different $z$ binnings:
$\Delta z = 0.005, 0.010,$ and $0.040$, shown as a function of the
galaxy bias.  The four lines in each panel correspond to halo
occupation slopes of $S=1.1, 1.0, 0.8, 0.6$ and $0.4$.  The number
density of galaxies is fixed to the observed value for
spectroscopically-confirmed LBGs.  The sharp upturn in the $S=0.4$
line occurs at a bias of $\sim 3.2$, which is the maximum bias for an
$S=0.0$ model ($p=1.0$).  In order to obtain a bias greater than this
for the $S=0.4$ case, $\Mmin$ must be larger than $\Mn$, which forces
all haloes to host more than one galaxy on average, and thus increases
the close pair fraction drastically.  The data points and solid-line
error bars show determinations of $f_{cp}$ and $b_g$ from the A01 and
Adelberger (2000) samples respectively.  The dotted-line error bars
reflect the range in $b_g$ determinations from a number of recent
estimates (see text), providing a reasonable estimate of the
systematic uncertainty in determining $b_g$.  }
\label{fig:noscat}
\end{figure*}

The thin vertical line corresponds to an $S=0.0$ model with $p=1.0$.
This corresponds to the simple `one-galaxy-per-massive-halo' type
model that has often been considered in previous works.  As we
discussed in Section~\ref{sec:observations}, for the special case
$S=0$, $\Mn$ is irrelevant and we must assign a value for the
selection probability $p$ in order to obtain a value for $\Mmin$ that
provides the desired (observed) density $n_g$.  
Since $p$ cannot exceed $1.0$, this vertical line represents an upper
limit on the value of $\Mmin$ for a given value of
$n_g$\footnote{Although it should be noted that for the LBG sample,
the known selection effects due to sub-sampling and the
color-selection technique mean that the selection probability is
almost certainly less than about 25 percent ($p \lsim 0.25$); see
Section~\ref{sec:observations}. 
Taking this into account would lead to a correspondingly lower value
for $\Mmin$. }.  Note that for this reason, it is possible to find
combinations of $n_g$ and $b_g$ that cannot be reproduced
simultaneously in this sort of model (ie. if $n_g$ is `too high' for
the given bias value). Similarly, as $p$ is lowered in an $S=0.0$
model, subject to the constraint that the \textit{observed} value of
$n_g$ is fixed, the implied `true' number density increases, and
$\Mmin$ must be reduced accordingly.  Since low-mass haloes are less
strongly clustered, the corresponding predicted bias will also
decrease.  The (dashed) lines of constant bias in Fig.
\ref{fig:m0-mmin} approach vertical asymptotes at the value of $\Mmin$
that produces this bias in the $S=0.0$ case. For example, if $S=0.0$
and $p=0.1$, then the observed number density requires $\Mmin \simeq
1.2 \times 10^{11} \hMsun$ and gives $b_g \simeq 2.2$.  Note that the
$b_g = 2.2$ dashed line approaches its vertical asymptote at the same
value of $\Mmin$.

Even if only the number density and large-scale bias are known,
Fig.~\ref{fig:m0-mmin} already places strong limits on the halo
occupation function for $z\sim3$ galaxies.  For example if the
observed LBG bias is constrained at $b_g < 3$, then the critical mass
above which halos host more than one observed galaxy at $z \sim 3$ must be
rather large, $\Mn \ga 4\times 10^{12} \hMsun$, regardless of the
values of $S$ and $\Mmin$.  If $b_g$ is estimated to lie within some
well-defined range, then, independent of the the occupation function
slope $S$, the model parameters $\Mn$ and $\Mmin$ must lie in a region
defined by the two (dashed) bias lines corresponding to that range.
For example, the bias estimate from Adelberger (2000) is $b_g \simeq
2.2-2.5$.  The allowed region implied by this measurement is indicated
by the shaded band in Fig.~\ref{fig:m0-mmin}.  Specifically, for $\Mn
\ga 5 \times 10^{13} \hMsun$ this implies $\Mmin \simeq (1-2)\times
10^{11} \hMsun$, and for $\Mmin \la 10^{11} \hMsun$ (and $S > 0$),
it implies $\Mn \simeq (5-50)\times 10^{12} \hMsun$.

In Fig. \ref{fig:noscat}, we have again fixed the number density to
match the observed value, but now we plot the model predictions in the
plane of close pair fraction versus the large-scale bias (note that
both quantities are directly observable).  Recall that in estimating
$f_{cp}$, we use angular bins of radius $20''$ ($0.4$ comoving \hmpc\
at $z=3$ in our adopted cosmology).  The pairs are defined in redshift
bins of $\Delta z = 0.005, 0.010,$ and $0.040$, as indicated
separately in each panel.  The solid lines show model predictions for
fixed values of the slope, $S= 1.1, 1.0, 0.8, 0.6,$ and $0.4$.  The
high-$S$ lines lie above those of lower $S$.  The lines are truncated
at a bias corresponding to $\Mm = 10^{8} \hMsun$.  This is an
extremely conservative lower limit on the expected mass of halos that
can host LBGs --- two orders of magnitudes smaller than lower limit on
the LBG host mass derived by \cite{pettini:01} using equivalent widths
of nebular emission lines.

The curves tend to be flatter at low bias and to rise more steeply at
large bias.  This corresponds to a transition between a close pair
density that is dominated by projection ($n_{\rm cp}^{\rm d}$) at low
bias, and is dominated by objects within the same halo ($n_{\rm
cp}^{\rm h}$) at high bias (see Eq. \ref{eq:ncp}).  As the close pair
fraction is calculated for larger and larger bins in redshift $\Delta
z$, the contribution due to galaxies in projection becomes larger,
while that coming from objects within the same halo remains the same.
This is why, for low-$b_g$, the values of $f_{cp}$ generally change
from one panel to the next, while for high-$b_g$ there is very little
change.  This tendency also provides an additional constraint.  In
principle the model can be constrained with one choice of $\Delta z$,
but by looking at data with various choices of the redshift bin, we
can get an additional handle on the fraction of the close pairs coming
from same-halo galaxies and from projection effects.  For example, a
strong change in close pair fraction with $z$ binning would indicate
that the close pairs are dominated by projection effects.

The behavior at high bias, corresponding to the single-halo dominated
regime, can be understood by examining Fig. \ref{fig:m0-mmin}.  For
a fixed slope, the typical number of objects within the same halo
depends primarily on the value of $\Mn$ relative to $\Mmin$.  Since
$\Mn$ is the mass above which haloes typically host more than one
galaxy, if $\Mn$ is much larger than $\Mmin$, then most haloes will
have fewer than one object, and the close pair fraction will be small.
Similarly, if $\Mn \simeq \Mmin$ then a large fraction of haloes will
host multiple galaxies, and the close pair fraction is high.  The
slopes of the $f_{\rm cp}-b_g$ relation are steeper for low-$S$ models
because the $\Mn-\Mmin$ relation is also steeper for these models.
The transition from projection-dominated to same-halo-dominated occurs
at larger bias for lower $S$ models for the same reason.  In contrast
to the single-halo piece, the projection-dominated (low-bias) regime
tends to be steeper for high-$S$ models.  The reason again is
associated with the slopes of the $\Mn-\Mmin$ curves shown in Fig.
\ref{fig:m0-mmin}.  The projected close pair density, $n_{\rm cp}^{\rm
d}$, (Eq. \ref{eq:ncpd}) is proportional to the amplitude of the
distinct halo correlation function.  Therefore one might expect that
$n_{\rm cp}^{\rm d}$ simply would grow as a function of bias, and
indeed it does for high-$S$ models.  The reason why the low-$S$ models
show very little change as a function of bias is that this tendency is
compensated by the rapid change in $\Mmin$ as a function of bias in
these models.  As $\Mmin$ increases, so does the typical halo size,
and therefore the region over which the halo-exclusion drives the
separate halo correlation function to zero becomes larger.  For
$S=0.4$, the typical halo exclusion size grows so rapidly with the
bias that it cancels out the effect of increasing the correlation
function amplitude at large scales. (The $S=0$ case looks nearly
identical to the $S=0.4$ out to a bias of about $3.2$, which is why we
do not plot any smaller values of $S$).

The data points on Fig. \ref{fig:noscat} show the estimates of close
pair fraction and $1-\sigma$ uncertainties derived from the A01 data
set, as discussed in Section~\ref{sec:observations}, along with the
bias value and its statistical $1-\sigma$ uncertainty from 
\citet{adel:00}.  We use the bias and 
error from \citet{adel:00} because this sample is very similar to that
of A01.  However, the formal  uncertainty quoted for this value likely
underestimates  the  precision to which  we know  the  value of $b_g$,
since  best estimates tend to  vary for  different samples and methods
(see Table 1).  In order to allow for this, we show a dotted error bar,  
which spans  the range  of  recent determinations  ($b_g =
2.1-2.9$).  From now on,  we refer  to the \citet{adel:00} uncertainty
(solid line   error bar) as  our  fiducial bias  range  and the larger
uncertainty (dotted line error bar) as our our expanded range. 

The data seem  to favor a  model with $S \simeq  1.0$.  The  fact that
this slope  seems to  reproduce the  observed $f_{cp}$  counts (within
fiducial errors) for all three $\Delta z$ binnings is encouraging, and
is   also an indication  that  our  simplified  halo occupation  model
provides  a reasonable description  of the galaxy-to-halo relation for
this sample.    The $\Delta z   = 0.01$ panel  provides  the strongest
constraint,  and within the   fiducial $1-\sigma$ uncertainties shown,
the compatible range of model slopes is $S\simeq 0.9-1.1$, with larger
values of bias preferring  shallower slopes.  The expanded bias  range
is consistent with a slightly larger slope range $S \simeq 0.8-1.1$.

\begin{figure*}
\resizebox{0.47\textwidth}{!}{\includegraphics{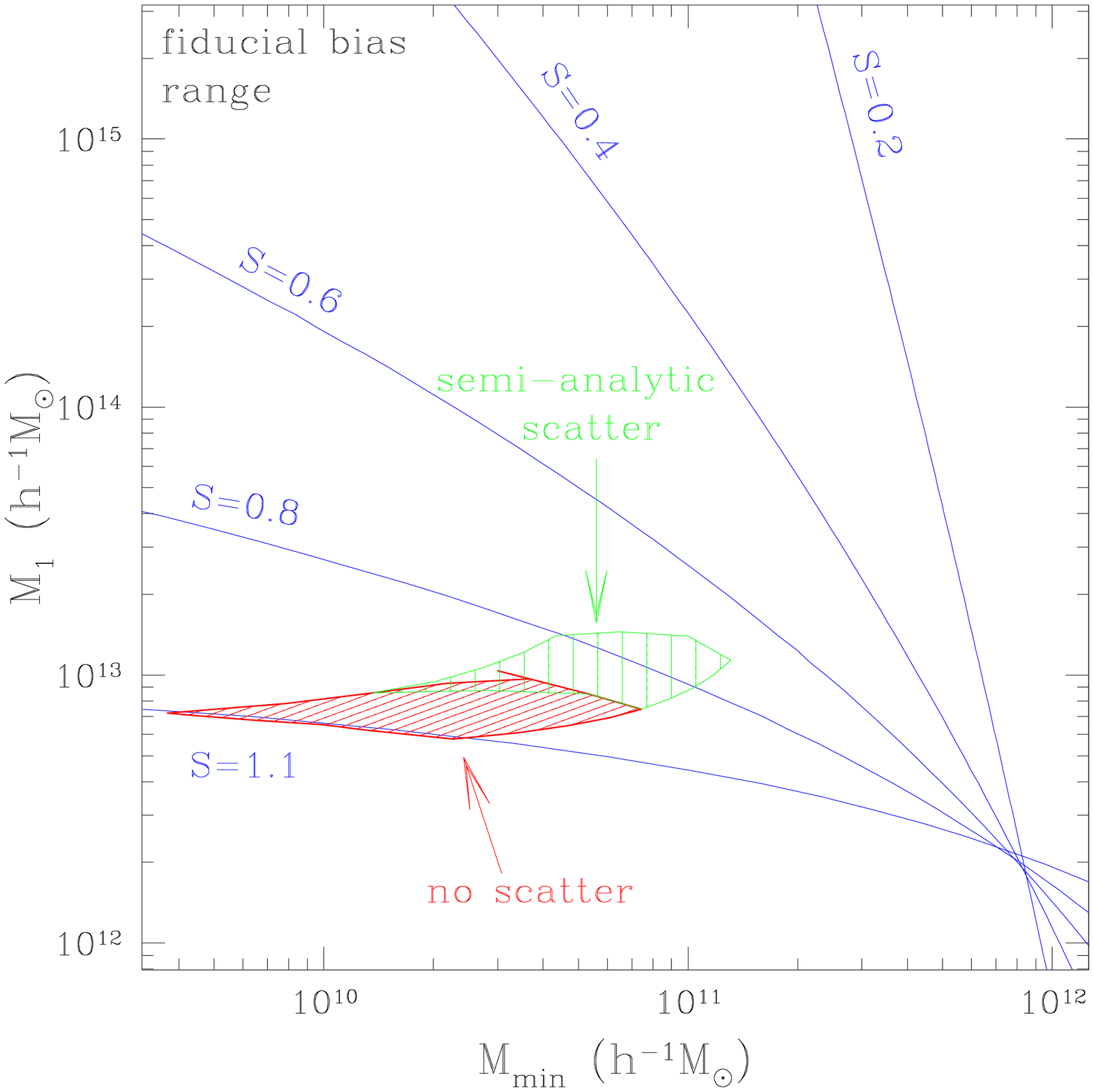}}
\resizebox{0.47\textwidth}{!}{\includegraphics{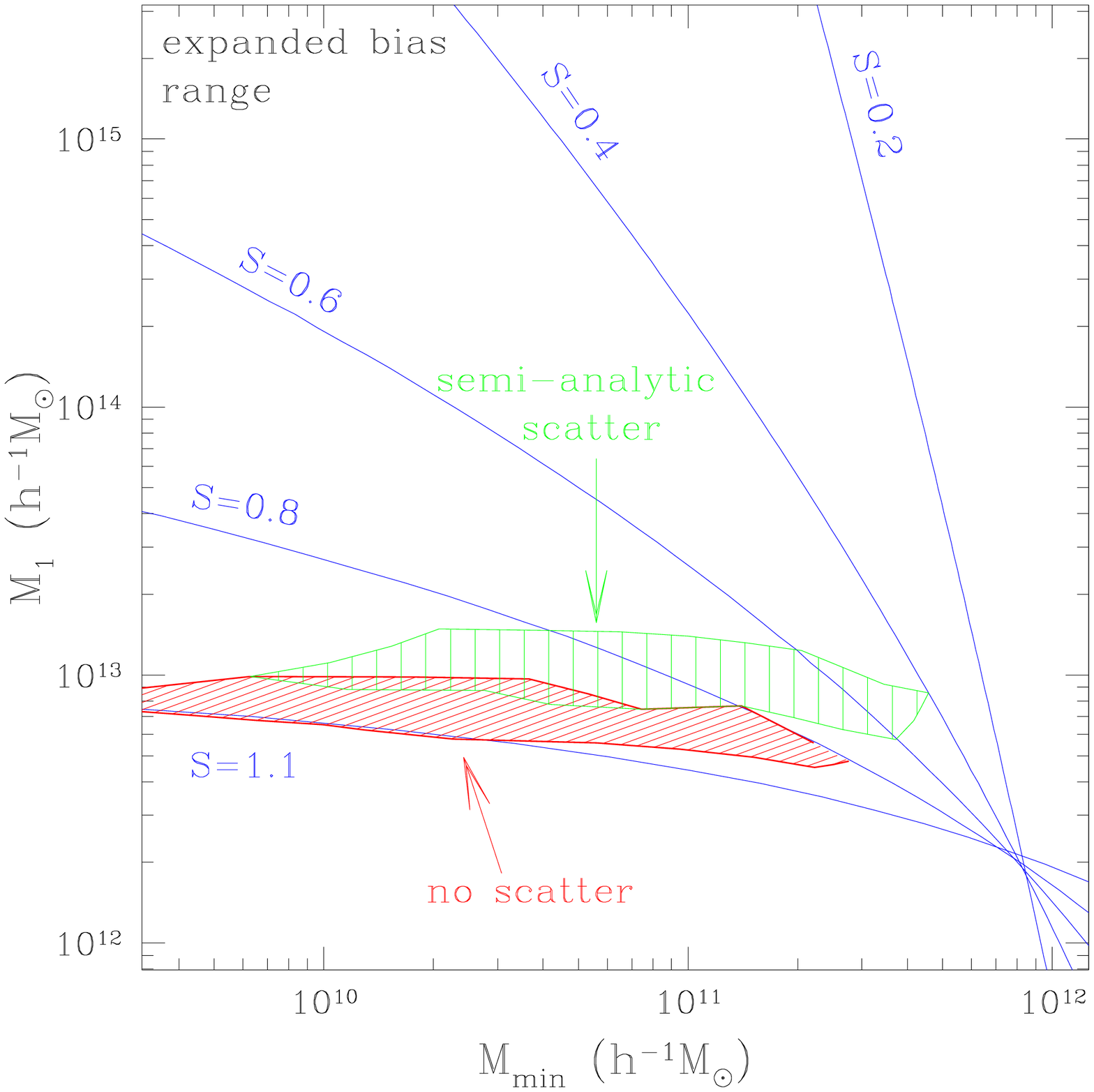}}
\caption[Allowed model parameters] {Allowed region of halo occupation
function parameter space implied by all three observational
constraints ($n_g$, $f_{cp}$ and $b_g$), with  and without scatter in
the halo occupation statistics.  Solid lines show models of
constant slope $S$.  (Left) Filled 
regions indicate the parameter space consistent with $1-\sigma$ bias
errors from Adelberger (2000).  (Right) Filled regions 
show how the allowed model space expands when a larger range for the bias
is allowed (see text).}
\label{fig:allowed}
\end{figure*}

Fig. ~\ref{fig:allowed} maps the constraints shown in Fig.
\ref{fig:noscat} to the allowed regions in $\Mn$ and $\Mm$ parameter
space.  Allowed regions are filled with closely 
spaced slanted lines. 
For each observationally-consistent value of $b_g$, there is a
range of  $S$ values that  are  allowed by  the close pair constraint.
Correspondingly,  there are well-defined  regions  in $\Mn$ and  $\Mm$
space consistent with each $S$  and $b_g$ combination, as indicated by
the  slanted line-filled bands.  It is important to realize that
each point in a filled region represents a unique model combination
of $S$, $\Mn$, and $\Mmin$.  The solid lines, representing models
of constant $S$ in this space, are shown to
help guide this understanding.

The left panel shows the region consistent with our fiducial
bias range and the right panel corresponds to the expanded bias
range.
No strong statistical significance should be attached to the
filled bands ---
they simply represent areas in parameter space that provide overlap
between the theoretical predictions and the data shown in
Fig. \ref{fig:noscat}. We have neglected any error in the analytic
estimates, which is likely of the order of $\sim 10$ per cent in
$f_{cp}$ and $b_g$.  The main point to take away is that the allowed
parameter space has been significantly reduced compared to that shown
in Fig.
\ref{fig:m0-mmin}.  For our fiducial bias uncertainty
we find that $\Mmin$ should lie roughly in the range
$(0.4-8)\times 10^{10} \hMsun$ and $\Mn \simeq (6-10) \times 10^{12}
\hMsun$.  For the expanded uncertainty range, the allowed 
space for $\Mmin$ expands as well,  $\Mmin \simeq (0.1 - 20) \times
10^{10} \hMsun$, but the $\Mn$ range remains roughly the same.  
Note that larger allowed values for $\Mmin$ correspond to
\textit{lower} values of $S$ (at fixed $f_{cp}$).

It is useful at this point to explore how an intrinsic scatter in halo
occupation would affect these estimates of the allowed model parameter
space.  Until this point,   we have assumed a  deterministic  relation
between a halo's  mass and the  number  of galaxies it  hosts, but any
realistic model of galaxy formation will  surely predict at least some
scatter  in this quantity.  In principle,  this distribution about the
mass-occupation  relation could be treated as  an  additional input of
the model,  however, here we will work  out a simple case motivated by
semi-analytic  galaxy formation scenarios in  order  to illustrate how
scatter affects our results.

Scatter  in the halo occupation  function will have  no  effect on the
predictions of  number  density and  large-scale  bias, and will alter
only the  close pair  fraction associated with   same-halo  pairs.  In
order to  account for  these,   we  must alter  Eq.~\ref{eqn:ncp}   by
replacing $\Ng(\Ng   -1)$  with    the  appropriate  expression    for
$\left<\Ng(\Ng    - 1)\right>$.  Although  one  might   suppose that a
Poisson distribution  would be a reasonable assumption, 
$\left<\Ng(\Ng - 1)\right> = \Ng^2(M)$, such an assumption 
is physically unrealistic for $N_g  \la 1$,  or $M  \la \Mn$.  As  the
host  mass falls  below that  typical  for containing  an  object, the
likelihood for  it to host  any additional objects  becomes suppressed
simply by mass counting  arguments.  This kind of  sub-Poisson scatter
is seen for low-mass hosts in semi-analytic models of galaxy formation
\citep[e.g.][]{scoc:01}.   For  our illustrative example,  we will use
the same   halo  pair counting  observed  in the  semi-analytic models
presented in W01 and
\citet{spf:00}, which becomes sub-Poisson below $\Ng \sim 1$:
\begin{eqnarray}
\left< N_{g}(N_{g}  - 1)\right> =  
\left\{ 
\begin{array}{cc}
	N_g^2  &   \mbox{ $\Ng \ge 1$} \\
	N_g^2 \ln(4 N_g)/\ln(4)   &  \mbox{$ 0.25<\Ng<1$}  \\        
	0 &  \mbox{ $\Ng \le 0.25$}
             \end{array}  
\right. 
\end{eqnarray}
We have suppressed the implicit mass dependence in $\Ng \equiv
\Ng(M)$.  Although the \citet{spf:00} models are best described by an
occupation function with $S \sim 0.7-0.8$, we assume here that the
above formula holds for all values of $S$.

\begin{figure*}
\centerline{\psfig{file= 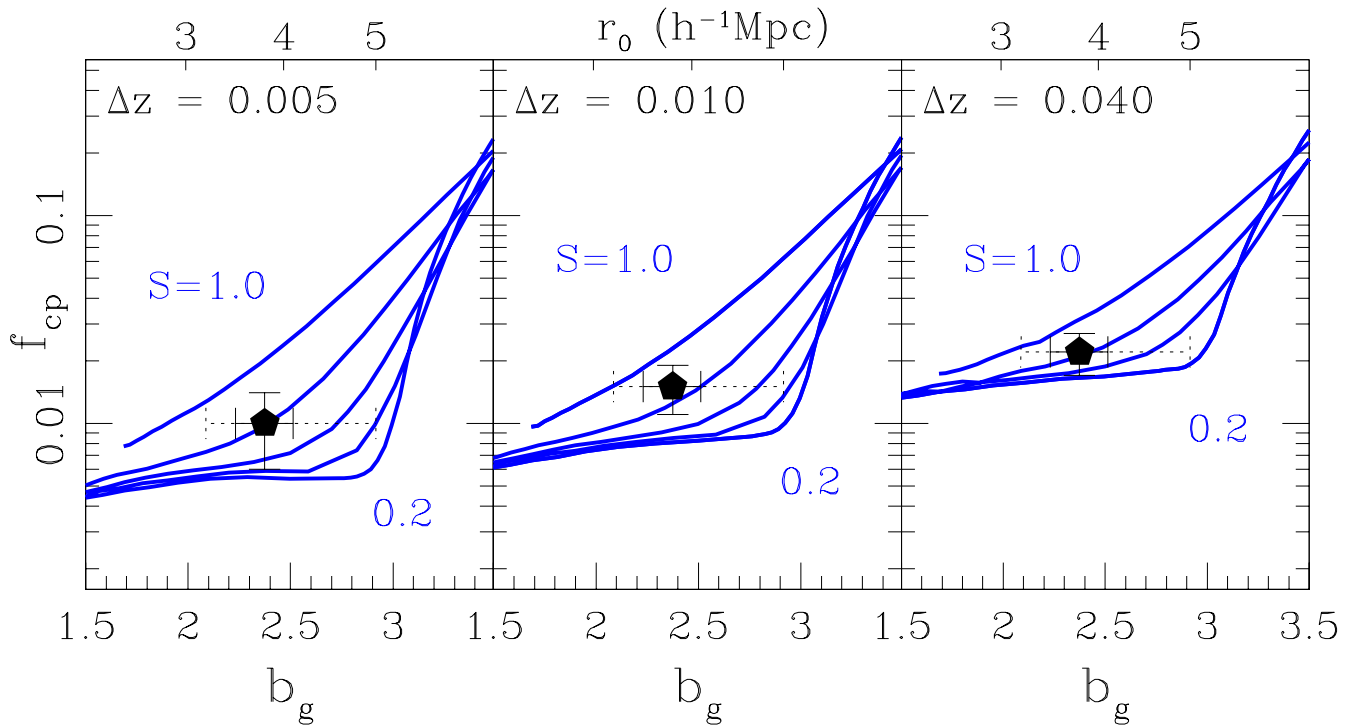,width=7in}}
\caption[Pair fraction vs. galaxy bias, for models including scatter]
{Close pair fraction as a function of bias for different values of the
halo occupation slope, as in Fig.~\protect\ref{fig:noscat}, but
including a non-zero scatter about the halo occupation relation, as
described in the text.  The number density is fixed to the observed
value for the A01 sample of Lyman-break galaxies, as usual.  The
fraction of pairs within an angular separation of 20 arcsec is shown for
three different $z$ binnings: $\Delta z = 0.005, 0.010,$ and $0.040$,
as a function of large-scale bias.  The four lines in each panel
correspond to halo occupation slopes of $S=1.0, 0.8, 0.6, 0.4, $ and
$0.2$.  }\label{fig:scat}
\end{figure*}

The results of  this calculation are  shown  in Fig.   \ref{fig:scat}.
When  the  average  number   of  objects  is small, including  scatter
increases the probability of having multiple objects in the same halo,
and  consequently increases  the expected close  pair  count. For this
reason, the steeply rising portions of the $f_{cp}-b_{g}$ curves begin
to become important  at lower  values of $b_g$  relative  to those  in
Fig. \ref{fig:noscat}.  It is encouraging that even with the inclusion
of this substantial amount  of  scatter, the  data  appear to favor  a
slope similar to  that suggested by the  zero-scatter models  shown in
Fig.  \ref{fig:noscat}, although slightly  lower, $S \simeq 0.9$.  The
fiducial $1-\sigma$ errors overlap with the model  curves for $0.7 \la
S   \la 1.0$, and the   expanded bias uncertainty  extends the allowed
range to  somewhat  shallower   slopes, $0.4 \la   S  \la  1.0$.   The
corresponding allowed regions  in  $\Mn$-$\Mm$ space for the  fiducial
(left panel)  and expanded (right panel) bias ranges are indicated by
the vertically-shaded bands in Fig.  \ref{fig:allowed}.  The ranges of
preferred $\Mm$  values have  significant overlap  with those  in  the
zero-scatter case:  $\Mmin   \simeq (1-13)\times 10^{10}   \hMsun$ for
fiducial  bias uncertainty, and  $\Mmin \simeq (0.6-40) \times 10^{10}
\hMsun$  for the expanded case.  However, the preferred  
regions are offset
towards higher $\Mn$ by roughly a
factor  of two; this is  due to the  fact that  increasing the scatter
increases the close  pair fraction and  thus the number of objects per
halo must be decreased by   increasing $\Mn$. Note, however, that  the
constraint   on the minimum   halo mass  $\Mm$ is  more  robust to the
inclusion of scatter.

\section{Luminosity/Number Density Dependence}
\label{sec:lumseg}
So far, we have considered observational constraints for a population
of galaxies with a given magnitude limit.  However, it is interesting
to consider how the predicted properties would change for samples with
different magnitude limits. Because the observed number density is a
function of the magnitude limit of the sample (for a sample of known
completeness), this can also be thought of as considering different
values for the observed number density.

Such a prediction is particularly relevant in light of the work by
\citet[][GD01]{giav:01}, which suggests a correlation between the LBG
clustering amplitude and intrinsic galaxy luminosity.  GD01 compared
correlation lengths obtained from ground-based spectroscopic and
photometric samples, and a deeper sample of LBGs identified in the
Hubble Deep Field.  They found that the correlation length strongly
decreased as the magnitude limit of the sample grew fainter, or,
similarly, as the observed number density of the population increased.
If correct, this result has interesting implications for the
relationship between observed galaxies and dark haloes.
Unfortunately, a prediction for the expected clustering as a function
of observed number density is rather unconstrained in our model.  This
is because, in principle, all three of our model parameters could vary
as a function of galaxy luminosity.

We can overcome this problem by making some plausible simplifying
assumptions.  For example, perhaps the simplest possibility is that
the value of $S$ stays fixed when the number density/luminosity cutoff
of a sample changes (which is roughly true in the semi-analytic models
of \citealt{spf:01}), and that $\Mmin$ and $\Mn$ vary together in the
natural way $\Mmin \propto \Mn$.  This assumption may be motivated
qualitatively by assuming that $\Mmin$ is proportional to the minimum
observable galaxy luminosity, and that the host haloes themselves are
self-similar as smaller and smaller haloes become important.  For
$S=0$, the $\Mn$ assumption cannot apply, and is replaced by the
assumption that selection probability $p$ remains fixed.

\begin{figure}
\PSbox{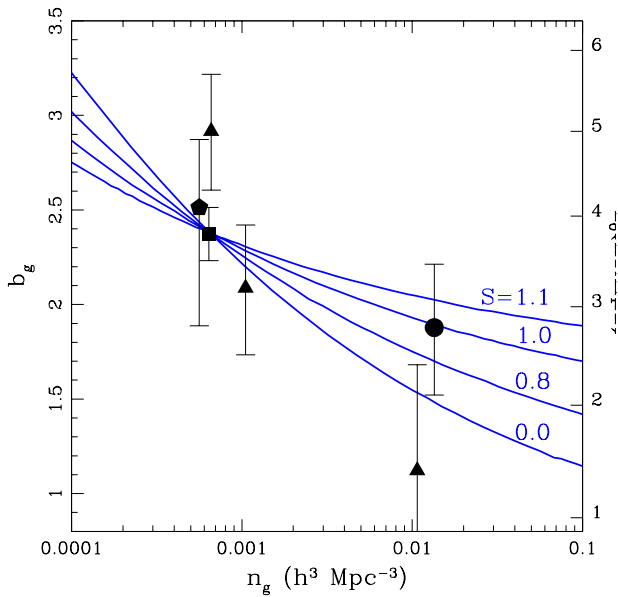  hoffset=-10 voffset=-65 hscale=40 vscale = 40}
{74mm}{74mm}
\caption[Bias vs. comoving number density] {Bias (left axis), and
correlation length (right axis) as a function of comoving number
density, for various values of the slope.  The data points are
described in the text.  Note that the points at high and low number
density have been offset slightly in the horizontal direction in order
to allow their error bars to remain distinct.}
\label{fig:dens}
\end{figure}

The resulting model predictions are shown in Fig. \ref{fig:dens},
for four values of the halo occupation slope $S$.  We have normalized
each model so that it has $b_g = 2.4$ at the number density of the
spectroscopically confirmed sample of \citet{adel:00}, and predict how
the bias should vary as function of $n_g$.  The $S=0$ model shows the
steepest dependence because the number density can be increased only
by adding galaxies to increasingly lower mass (and less clustered)
haloes.

The data points correspond to the observational estimates.  The
triangles show the results of GD01.  The square reflects
the \citet{adel:00} estimate.
The angular correlation
function result from recent work by \citet{porciani:01} is shown by
the pentagon.  The filled circle shows the results of
\citet{arnouts:99} based on a sample from the HDF with a similar
magnitude limit as the GD01 HDF data, but selected via
photometric redshifts rather than the Lyman-break technique. 
We have not included the \citet{adel:98} determination of the bias
because it has been superseded by the \citet{adel:00} sample, which
uses a larger sample of galaxies and the same counts-in-cells
method.
In light of the disagreement between the various estimates at fixed
density, the strength of the trend must be regarded as rather
uncertain.  If the GD01 points are neglected, then all four models are
consistent with the data, but taken together, the data seem to favor a
model closer to the $S=0.0$ line, unlike the close-pair and bias
constraints discussed in the previous sections, which favored $S\sim
0.9-1.0$.

However, it is important to note that the GD01 points cannot be
reproduced by any of these minimum-assumption models, as even the
$S=0.0$ model trend is too shallow.  The only way to obtain such a
trend would be to assume that LBGs tend to \emph{avoid} the most
massive haloes, corresponding to a negative slope $S$.  Otherwise a
trend between clustering and number density as steep that indicated by
GD01 can only be accounted for by breaking one of our simplifying
assumptions.
For example, if the selection probability, $p$, varies systematically
with the observed number density, then the low bias of the higher
density sample might indicate that the true number density is simply
much larger than that observed.  We find that only if the selection
probability varies inversely with the observed number density: $p
\propto n_g^{-1}$ (with $p \le 1$), can we reproduce a trend this
steep with $S=0$.  Although one might expect the selection probability
to be higher for brighter objects, such a strong trend seems
problematic for other reasons.
Recall that the value of $p$ can only effect the estimate of the
number density, not the bias, so this would imply that there is almost 
no difference in the number density of LBGs at ${\mathcal R} = 25$ and 
${\mathcal R \sim 27}$, which is at odds with observational
determination of the luminosity function of LBGs over this range
\citep{steidel:99}. 

\section{Conclusions}
\label{sec:conclusions}

We have presented  a simple and  intuitive method for constraining the
relationship  between observed galaxies  and their dark matter haloes,
and used it to constrain the galaxy-halo occupation relation at 
$z \sim 3$.  Using a three-parameter model of the form 
$N_g(M) =(M/\Mn)^S$,  $M > \Mmin$ to describe  the number of observed 
galaxies
as a  function the host  halo  mass, we derived  predictions for three
observables:  the  galaxy  number  density, $n_g$,   large-scale bias,
$b_g$, and  the fraction of galaxies  in close pairs, $f_{cp}$.  Given
these three    observed galaxy properties,   the   three unknown model
parameters can be constrained.

We presented estimates of the allowed range for these three parameters
based on the  properties of galaxies in  a sample of spectroscopically
confirmed LBGs \citep{adel:00, adel:01th}.  The results are summarized
in Fig. \ref{fig:allowed}. For a model  with no scatter about the halo
occupation relation, the favored values of the slope lie in the range 
$0.9 \la S \la  1.1$, with preferred  characteristic  halo masses
 $\Mm   \simeq (0.4-8)\times 10^{10}  \hMsun$  and 
$\Mn \simeq (6-10) \times 10^{12} \hMsun$.  
For a  model where the halo occupation
function scatter is estimated based on semi-analytic models, the range
of preferred  model  parameters shifts to  $0.7  \la S  \la 1.0$, 
$\Mm \simeq  (1-13)\times 10^{10}  \hMsun$  and 
$\Mn  \simeq (8-15)  \times10^{12} \hMsun$.  
Since the observational uncertainty in $b_g$ is likely significantly
larger than the formal error derived for the sample of LBGs we
consider, we have also explored a range of biases consistent
with all of the recent estimates for LBG clustering.  Using
this expanded range of uncertainties, we obtain
$0.8  \la S  \la 1.1$ and
$\Mm   \simeq (0.01-20)\times 10^{10}  \hMsun$
for the case  of no scatter, and
 $0.4  \la S  \la 1.0$, $\Mm   \simeq (0.6-40)\times 10^{10}  \hMsun$
if scatter is included.  Preferred values of $\Mn$ are relatively
insensitive to increasing the uncertainty in $b_g$, 
and preferred values of $\Mmin$ are relatively insensitive to the
inclusion of scatter in the occupation function.

It is interesting that the data favor a model in which the minimum
halo mass for hosting an observable object, $\Mmin$, is significantly
smaller (by roughly two orders of magnitude) than the mass range above
which haloes typically host more than one observed object, $\Mn$.
However, recall that the average mass halo hosting an {\em observable}
object is typically lower: $\Mn^i = p^{1/S}{\Mn}$, so the difference
between $\Mmin$ and $\Mn^i$ ranges from about two orders of magnitude for
high $S$ values to no difference for $S=0$.  
In any case, this implies that
\emph{most} haloes hosting LBGs do not contain more than one
observable object. However, as we have seen, because the most massive
haloes are also the most clustered, the clustering predictions are
quite sensitive to the treatment of occupation statistics in these
haloes even though they constitute a small fraction by number. In
addition, it is interesting that the range of allowed values for
$\Mmin$ have considerable overlap with mass estimates based on the
widths of nebular emission lines, $\sim 10^{10} \hMsun$
\citep{pettini:01}, bearing in
mind that these line-widths may yield underestimates of the true
virial masses.  Line-width analyses such as this may provide useful
additional constraints for the type of model presented here.  For
example, one might be inclined to eliminate models with $\Mmin \la
10^{10} \hMsun$ based on the \citet{pettini:01} analysis, and thus
significantly reduce the allowed model parameter space.  However,
these line widths estimates are based on a relatively small sample of
very bright objects, so placing strict limits on $\Mmin$, which is
generally much smaller than than a `typical' LBG host, $\Mn$, may
not be justified.

Although the current level of observational uncertainty prevents us
from precisely defining a favored model, the identification of an
allowed range of parameter space already provides useful constraints
on more sophisticated modelling aimed at understanding halo occupation
at a more basic level.  For example, as mentioned in the introduction,
the first, and, for some, still the favored model for LBG occupation
is one in which there is one galaxy per halo, $S=0.0$
\citep[e.g.][]{wech:98}.  Our results disfavor such a model because it
under-predicts the close pair counts (see also W01).  Similarly, naive
models for collision-driven LBGs \citep{kolatt:99} predicted a steep
halo occupation function, $S \simeq 1.1$, which is the steepest slope
consistent with our results and requires that LBGs populate a fraction
of very low-mass halos (although in W01 we explain why the more
sophisticated treatment of collisional bursts presented in
\citet{spf:01} yields a shallower
relation, $S \sim 0.7$).  The preferred range of slopes from our
analysis including scatter is in agreement with the range of slopes
from semi-analytic estimates for the occupation function for a range
of star formation models (W01), including those in which the primary
mode of star formation is merger-driven starbursts, and also those in
which quiescent star formation dominates. However, it seems unlikely
that the slope will be constrained well enough in the near future to
distinguish {\em between} the different semi-analytic models compared
in W01.  Work by \citet{porciani:01} using a counts-in-cells analysis
of the angular correlation function of LBGs (using an overlapping but
different sample of galaxies than considered here) favors a model with
a shallower slope; their analysis is consistent with S=0.  Some of the
discrepancy may result from the more highly biased sample they
consider.
(The full A01 sample includes new fields that are less clustered than
the first few fields studied).  As illustrated in Fig. 2, for a fixed
$f_{cp}$, a higher bias favors a shallower slope.  Nonetheless,
neither our constraints or theirs are very strong, and we emphasize as
they do that current samples may not yet be a fair representation of
the high-redshift galaxy population.  When the sample becomes larger,
these complementary types of analyses should be applied in parallel to
constrain the halo occupation function.

We then applied this simple model to investigate how the clustering of
a population of galaxies might change as a function of their observed
number density, or, implicitly, as a function of their intrinsic
luminosity.  Under the simplest assumptions for how model parameters
should vary as a function of luminosity cut, the low-$S$ models vary
more strongly with number density than do high-$S$ models.  None of
the models are steep enough to match the trend found by GD01, but
models with $S=0.0-1.0$ are consistent with the change in bias between
the sample of \citet{adel:00} and
\citet{arnouts:99} --- so this data cannot yet provide significant
constraints.  More data which can provide smaller error bars on
observational parameters, especially the bias, could prove a valuable
constraint for the halo occupation relation.  We have shown how
combinations of our three model parameters are currently constrained
by the data, but with the current data sample individual parameters
(such as the slope S or the minimum mass for hosting a halo $\Mmin$)
are not well determined.  This will probably await a significantly
larger survey, such as one that could be completed using, e.g., the
Very Large Telescope (VLT) or the Large Binocular Telescope (LBT).

We have discussed here how the formalism presented in \S
\ref{sec:model} can be applied to determine the halo occupation of a
specific population of LBGs at $z\sim3$, but in fact it is quite
general, and could be applied to constrain the halo occupation models
for a variety of galaxy populations, and used to understand the
relation between various galaxy populations.  The limitation of this
method is that, as we have shown, it requires a large sample to be
able to put strong constraints on model parameters.  Moreover, the
method relies on the simplifying assumption of scale-independent bias,
and in cases where the slope of the galaxy correlation function is
significantly different from that of the dark matter, it becomes
ill-defined.  However, with new generations of telescopes, large
samples will become available for an increasing number of types of
high redshift galaxies.  Once statistics become available from the
LALA survey \citep{rhoads:00}, a similar approach to the one we have
presented could be used to understand whether Lyman-$\alpha$ emitters
and LBGs populate the same dark haloes, or to relate the LBG population
to that of SCUBA sources --- such an analysis would be complementary
to the analysis of \citet*{shu:01} which uses star formation rates and
the observed size distribution to constrain halo occupation models.
Similar methods can also be used to relate the LBG population at $z=3$
with galaxy populations at different redshifts (Moustakas \&
Somerville, in preparation).  This basic formalism could also easily be
applied to quasars identified in the Sloan Digital Sky Survey, as a
tool for understanding the halo occupation of quasars as a function of
redshift and luminosity.

\section*{Acknowledgments}
We thank Kurt Adelberger for providing us with data on close pairs
and for help with interpreting the data.
Thanks also to Cristiano Porciani and Mauro Giavalisco for
helpful comments and providing us with an early draft of their results.  We
also thank Andreas Berlind, George Blumenthal, and David Weinberg for
a number of insightful conversations, and Joel Primack for 
useful comments on a draft.  JSB received support from NASA LTSA grant
NAG5-3525 and NSF grant AST-9802568, and RHW was supported by a DOE
GAANN fellowship at UCSC.

\bibliographystyle{mn2e}
\bibliography{risa}
\end{document}